\documentclass[10pt,a4paper]{article}
\usepackage{graphicx, array, blindtext}
\usepackage{amsmath}

\begin{document}			

\title{\Large{A Dark Energy Model interacting with Dark Matter described by an effective EoS.}}
\author{\small{Martiros Khurshudyan}\\
\texttt{\small{CNR NANO Research Center S3, Via Campi 213a, 41125 Modena MO, Italy}}\\\texttt{\small{and}}\\
\texttt{\small{Dipartimento di Scienze Fisiche, Informatiche e Matematiche,}}\\
\texttt{\small{ Universita degli Studi di Modena e Reggio Emilia, Modena, Italy}}\\ \\
\texttt{\small{email:martiros.khurshudyan@nano.cnr.it}}}
\date{\small{\today}}
\maketitle
\vspace{10 mm}
\begin{abstract}
In this latter author would like to consider interaction between a dark energy based on Generalized Uncertainty Principle (GUP) and a Dark Matter described by effective EoS: $P = (\gamma-1)\rho+p_{0}+\omega_{H}H+\omega_{H2}H^{2}+\omega_{dH}\dot{H}$ \cite{Martiros}-\cite{Ren}, which could be interpreted as a modification concerning to the some interaction between fluid $P=(\gamma-1)\rho$ with different components of the Darkness of the Universe. Two types of interaction, called sign-changeable, $Q=q(3Hb\rho_{m}+\beta\dot{\rho}_{m})$ \cite{Hao}, \cite{Martiros2} and $Q=3Hb\rho_{m}+\beta\dot{\rho}_{m}$ are considered. EoS parameter of the mixture $\omega_{tot}$ are investigated. Statefinder diagnostics provided also.
\end{abstract}
\newpage
\section*{\large{Introduction}}
\begin{figure}[htp!]
\centering
\includegraphics[width=60mm]{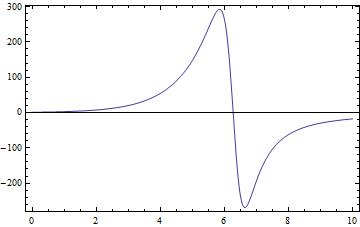}
\caption{$\ddot{a}$ against t}
\label{fig:newscale}
\end{figure}

\begin{figure}[htp!]
\centering
\includegraphics[width=60mm]{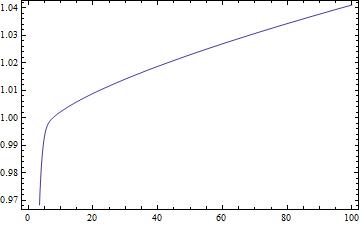}
\caption{Conformal time $\eta$ against t}
\label{fig:conftime}
\end{figure}

Statefinder diagnostics for Emergent, Intermediate and Logamediate scenarios showed, that  $Q=3Hb\rho_{m}$ form of interaction between a barotropic fluid and a dark energy model, based on Generalized Uncertainty Principle (origins of GUP are in the string theory), makes possible to cross $\{r=1,s=0\},$ corresponding to $\Lambda$CDM \cite{Ujjal}\footnote{Readers kindly requested to observe references of the article for self consistent information}. In this article we consider interaction of GUP Dark energy with a Dark Matter described by
\begin{equation}\label{eq:EoSp}
P = (\gamma-1)\rho+p_{0}+\omega_{H}H+\omega_{H2}H^{2}+\omega_{dH}\dot{H}
\end{equation}
EoS \cite{Martiros}-\cite{Ren}. We suppose, that such modification could be rise as a result of an interaction between $P=(\gamma-1)\rho$ fluid and Darkness of the Universe. Having this modification, we suppose that there is an interaction between GUP and modified fluid. Two different forms for interaction we take into account, one of them is called sign-changeable and includes deceleration parameter in order to provide sign changeability of the interaction term during evolution of the Universe, and the second type is interaction of ordinary form, which has the same sign from the early epoch to late stages. First type of interaction is described by $Q=q(3Hb\rho_{m}+\beta\dot{\rho}_{m})$ \cite{Hao}, \cite{Martiros2} and the second type reads as $Q=3Hb\rho_{m}+\beta\dot{\rho}_{m}$. For both cases we assume that interaction depends on matter energy density and its first order derivative only. Questions concerning to the nature of interactions term, appearing in cosmology just due to mathematics, are still open. Dark energy density based on GUP, which is described by (n, $\xi$) parameters, takes the form
\begin{equation}\label{eq:darkendens}
\rho_{G}=\frac{3n^{2}m^{2}_{p}}{\eta^{2}}+\frac{3\xi^{2}}{\eta^{4}},
\end{equation}
where $\eta$ is the conformal time and reads as
\begin{equation}\label{eq:conftime}
\eta = \int{\frac{dt}{a(t)}}.
\end{equation}
Hereafter, in the main writing of the article we will present results, concerning to Logamediate scenario: $a(t)=\exp{(\mu(\ln{t})^{\alpha})}$, with $\mu\alpha>0, \alpha>1$. In appendix we will present results for $a(t)=a_{0}(B+e^{At})^{m}$, with $a_{0}>0,A>0,B>0,m>1$ known as Emergent scenario and $a(t)=exp(\lambda t^{\beta})$, with $\lambda >0, 0<\beta<1$ called Intermediate scenario. We also perform analysis for a scale factor \cite{Kremer}, for which $\ddot{a}$ evolution over time presented in Fig.\ref{fig:newscale}, with such scale factor conformal time has a behavior shown in Fig.\ref{fig:conftime}.
\\ \\
The parameters of our interests are
the jerk or statefinder parameter
\begin{equation}\label{eq:statefinder}
r=\frac{1}{H^{3}}\frac{\dddot{a}}{a} ~~~~~~~~~~~~~ s=\frac{r-1}{3(q-\frac{1}{2})}.
\end{equation}
and $\omega_{\small{tot}}$ EoS parameter of the mixture.\\ \\
The paper organized as follow: in the next section we will introduce equations which governs our model. Next we will obtain and investigate parameters of our model for each type of interaction. At the end in the last section we will conclude results concerning to the model. In appendix results concerning to Emergent and Intermediate scenarios presented in order.

\section*{\large{The field equations and the Model}}
Field equations that governs our model read as
\begin{equation}\label{eq:Einstein eq}
R^{\mu\nu}-\frac{1}{2}g^{\mu\nu}R^{\alpha}_{\alpha}=T^{\mu\nu},
\end{equation}
which by means of FRW metric
\begin{equation}\label{eq:FRW metric}
ds^{2}=dt^{2}-a(t)^{2}\left(dr^{2}+r^{2}d\theta^{2}+r^{2}\sin^{2}\theta d\phi^{2}\right),
\end{equation}
is reduced to
\begin{equation}\label{eq: Fridmman vlambda}
H^{2}=\frac{\dot{a}^{2}}{a^{2}}=\frac{\rho_{\small{tot}}}{3},
\end{equation}
\begin{equation}\label{eq:Freidmann2}
- \frac{\ddot{a}}{a}=\frac{1}{6}(\rho_{\small{tot}}+P_{\small{tot}}),
\end{equation}
with Bianchi identities implying that
\begin{equation}\label{eq:Bianchi eq}
\dot{\rho}_{\small{tot}}+3\frac{\dot{a}}{a}(\rho_{\small{tot}}+P_{\small{tot}})=0.
\end{equation}
The mixture of our consideration describes by
\begin{equation}\label{eq:mixture energy}
\rho_{ \small{tot}} = \rho_{ \small{m}}+\rho_{ \small{G}}.
\end{equation}
\begin{equation}\label{eq:mixture presure}
P_{ \small{tot}} = P_{ \small{m}}+P_{ \small{G}}.
\end{equation}
In case of interaction between components, (\ref{eq:Bianchi eq}) splits into two following equations
\begin{equation}\label{eq:inteqm}
\dot{\rho}_{m}+3H(\rho_{m}+P_{m})=-Q
\end{equation}
and
\begin{equation}\label{eq:inteqG}
\dot{\rho}_{\small{G}}+3H(\rho_{\small{G}}+P_{\small{G}})=Q,
\end{equation}
where Q is the interaction term introduced above.
(\ref{eq:inteqm}) with (\ref{eq:EoSp}) allows us to obtain $\rho_{m}$. With $\rho_{m}$ and $\rho_{G}$ from (\ref{eq: Fridmman vlambda}) Hubble parameter as a function of $t$ accounting interaction under consideration will be obtained. Having $\rho_{m}$, $\rho_{G}$ and $H(t)$ from (\ref{eq:Freidmann2}) for $P_{G}$ in case of interaction will be covered immediately
\begin{equation}\label{eq:darkenergypressure}
P_{G}=-2\dot{H}-(P_{m}+\rho_{m})-\rho_{G}.
\end{equation}
Then having all parameters calculated we will be able to investigate parameters discussed above.
\section*{\large{Interaction $Q=3bH\rho_{m}+\beta\dot{\rho}_{m}$}}
\begin{figure}[htp!]
\centering
\includegraphics[width=60mm]{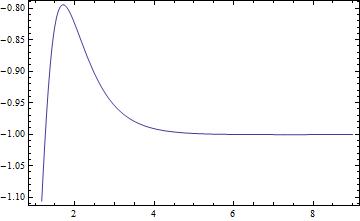}
\caption{Logamediate Scenario: $\omega_{\small{tot}}$ against t,
Interaction: $Q=3bH\rho_{m}+\gamma\dot{\rho}_{m}$}
\label{fig:omegatotLog}
\end{figure}

\begin{figure}[htp!]
\centering
\includegraphics[width=60mm]{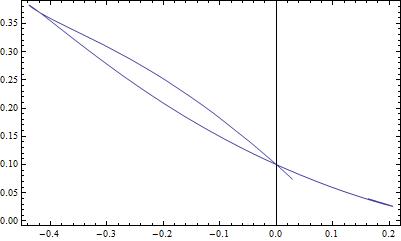}
\caption{Logamediate Scenario: $r\times10^{-1}$ against s,
$Q=3bH\rho_{m}+\gamma\dot{\rho}_{m}$}
\label{fig:rsLog}
\end{figure}

In Logamediate scenario for conformal time we have
\begin{equation}\label{eq:conformal3}
\eta=\int \frac{dt}{\exp(\mu(\ln t)^{\alpha})}.
\end{equation}
For $\alpha=2$ (\ref{eq:conformal3}) reads as
\begin{equation}\label{eq:conformaltime2}
\eta=\frac{e^{\frac{1}{4\mu}}\sqrt{\pi}Erf[\frac{-1+2\mu\ln t}{2\sqrt{\mu}}]}{2\sqrt{\mu}},
\end{equation}
where
\begin{equation}\label{eq:erf}
Erf(z)=\frac{2}{\sqrt{\pi}}\int_{0}^{z}{e^{-t^{2}}~dt}.
\end{equation}
For simplicity we omit expressions of Dark Matter Energy Density, of pressure and etc., and going to graphical analysis of the model. We consider $\alpha=2$ case in order to deal with analytical solutions, otherwise numerical analysis can be performed easily.
We start with $\omega_{\small{tot}}$ (Fig. \ref{fig:omegatotLog}). We observe that in early epoch $\omega_{\small{tot}}<-1$ and indicates phantom-like behavior, then during evolution the change of behavior from phantom to $\omega_{\small{tot}}>-1$ quintessence-like behavior occurs, which in its turn at late stages of evolution behaves as a cosmological constant with $\omega_{\small{tot}}=-1$. Statefinder diagnostics reveal that we cross $\{ r=1, s=0\}$ twice corresponding to $\Lambda$CDM. Departing from the section with positive $s$ and positive $r$ corresponding to the radiation phase of the Universe we cross $\{ r=1, s=0\}$ corresponding to $\Lambda$CDM and we back to radiation phase for late stage of the Universe (Fig. \ref{fig:rsLog}). This depends on a value of $\beta$ parameter presenting in interaction $Q$, which indicates an impact of the $\dot{\rho}_{m}$ in interaction \footnote{ $\mu=2.2$, $b=0.5$, $\beta=0.2$, $\gamma=1.5$, $p_{0}=-0.5$, $\omega_{H}=-1.5$, $\omega_{H2}=-0.6$, $\omega_{dH}=-0.3$ }.
\newpage
\section*{\large{Interaction $Q=q(3bH\rho_{m}+\gamma\dot{\rho}_{m})$}}
\begin{figure}[htp!]
\centering
\includegraphics[width=60mm]{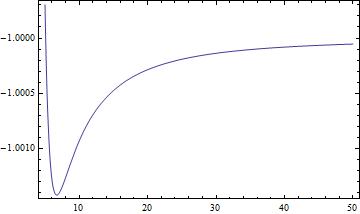}
\caption{Logamediate Scenario: $\omega_{\small{tot}}$ against t, Interaction: $Q=q(3bH\rho_{m}+\gamma\dot{\rho}_{m})$}
\label{fig:omegatotLog2}
\end{figure}

\begin{figure}[htp!]
\centering
\includegraphics[width=60mm]{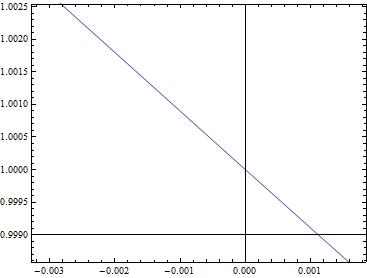}
\caption{Logamediate Scenario: r against s, Interaction: $Q=q(3bH\rho_{m}+\gamma\dot{\rho}_{m})$}
\label{fig:rsLog2}
\end{figure}

In this section we will consider sign-changeable interaction between components, which involves deceleration parameter $q$. Here we present graphical analysis of some parameters of the model. Parameters of our interest were found by the same mathematical way as in previous case. For EoS $\omega_{\small{tot}}$ parameter we found that in early stages of evolution it has quintessence-like behavior with $\omega_{\small{tot}}>-1$, then there is a transition to $\omega_{\small{tot}}<-1$ with phantom-like behavior and preserving it up to late stages of evolution (Fig \ref{fig:omegatotLog2}). Statefinder diagnostics presented in Fig.\ref{fig:rsLog2} \footnote{ $\mu=2.2$, $b=0.5$, $\beta=0.2$, $\gamma=1.5$, $p_{0}=-0.5$, $\omega_{H}=-1.5$, $\omega_{H2}=-0.6$, $\omega_{dH}=-0.3$ }.
\section*{\large{Scale factor 2}}
\begin{figure}[htp!]
\centering
\includegraphics[width=60mm]{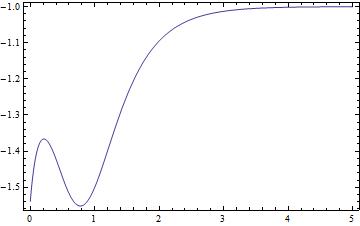}
\caption{Scale factor 2. $\omega_{\small{tot}}$ against t. Interaction: $Q=3bH\rho_{m}+\gamma\dot{\rho}_{m}$, $p_{0}=-1.5$}
\label{fig:omegatot2}
\end{figure}

\begin{figure}[htp!]
\centering
\includegraphics[width=60mm]{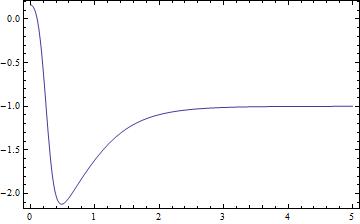}
\caption{Scale factor 2. $\omega_{\small{tot}}$ against t. Interaction: $Q=3bH\rho_{m}+\gamma\dot{\rho}_{m}$, $p_{0}=1.5$}
\label{fig:omegatot2_1}
\end{figure}

\begin{figure}[htp!]
\centering
\includegraphics[width=60mm]{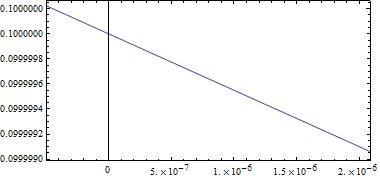}
\caption{Scale factor 2. $r\times10^{-1}$ against s, Interaction: $Q=3bH\rho_{m}+\gamma\dot{\rho}_{m}$}
\label{fig:rs2}
\end{figure}

\begin{figure}[htp!]
\centering
\includegraphics[width=60mm]{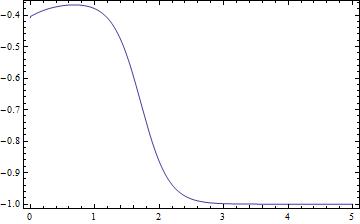}
\caption{Scale factor 2. $\omega_{\small{tot}}$ against t. Interaction: $Q=q(3bH\rho_{m}+\gamma\dot{\rho}_{m})$}
\label{fig:omegatot3_0}
\end{figure}

\begin{figure}[htp!]
\centering
\includegraphics[width=60mm]{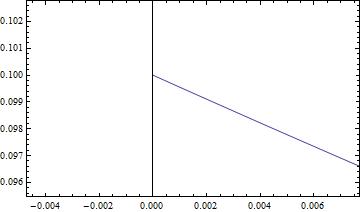}
\caption{Scale factor 2. $r\times10^{-1}$ against s. Interaction: $Q=q(3bH\rho_{m}+\gamma\dot{\rho}_{m})$}
\label{fig:rst3_0}
\end{figure}

\begin{figure}[htp!]
\centering
\includegraphics[width=60mm]{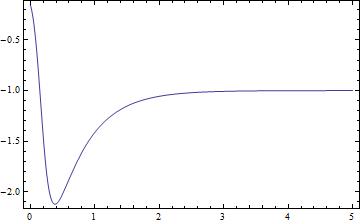}
\caption{Scale factor 2. $\omega_{\small{tot}}$ against t. Interaction: $Q=q(3bH\rho_{m}+\gamma\dot{\rho}_{m})$, $p_{0}=1.5$}
\label{fig:omegatot3}
\end{figure}

\begin{figure}[htp!]
\centering
\includegraphics[width=60mm]{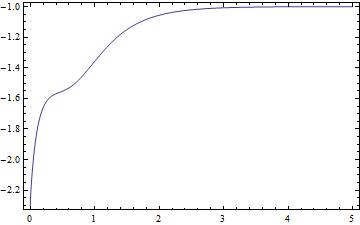}
\caption{Scale factor 2. $\omega_{\small{tot}}$ against t. Interaction: $Q=q(3bH\rho_{m}+\gamma\dot{\rho}_{m})$, $p_{0}=-1.5$}
\label{fig:omegatot3_1}
\end{figure}

Having scale factor profile for which $\ddot{a}$ over time is presented in Fig.\ref{fig:newscale} we perform numerical analysis in order to reveal the behavior of the parameters under consideration. With $Q=3bH\rho_{m}+\beta\dot{\rho}_{m}$ interaction we observe that when $p_{0}=\omega_{H}=\omega_{dH}=0$ and $\omega_{H2}=0,$ during whole evolution of the Universe $\omega_{\small{tot}}=-1$. Then for instance, when $p_{0}=-1.5,~ \omega_{H}=-1~ \omega_{dH}=-0.5$ and $\omega_{H2}=-0.65$ we observe that, $\omega_{\small{tot}}<=-1$ and it is true during whole evolution (Fig.\ref{fig:omegatot2}). In the case of positive $p_{0}=1.5$ we observe that in the early epoch $\omega_{\small{tot}}>0$ then there is stage with $\omega_{\small{tot}}>-1$ indicating quintessence-like behavior. Finally for late time evolution $\omega_{\small{tot}}<=-1$ indicates phantom-like behavior (Fig.\ref{fig:omegatot2_1}). Statefinder diagnostics shows that a crossing of $\{ r=1, s=0\}$ is possible (Fig.\ref{fig:rs2}). Taking into account sign-changeable interaction between components of the mixture we observe that in case $p_{0}=\omega_{H}=\omega_{dH}=0$ and $\omega_{H2}=0$ for EoS parameter we have $\omega_{\small{tot}}>=-1$ during evolution (Fig.\ref{fig:omegatot3_0}). For this case statefinder diagnostics presented in Fig.\ref{fig:rst3_0}. In case when parameters differ from $0$ we observe that with $p_{0}=1.5$ Universe starts with $\omega_{\small{tot}}>-1$, then continues its evolution and becomes $\omega_{\small{tot}}<-1$. Finally, for late time of the evolution $\omega_{\small{tot}}=-1$ (Fig.\ref{fig:omegatot3}). With $p_{0}=-1.5$ for EoS parameter we have Fig.\ref{fig:omegatot3_1}.
\begin{figure}[htp!]
\centering
\includegraphics[width=60mm]{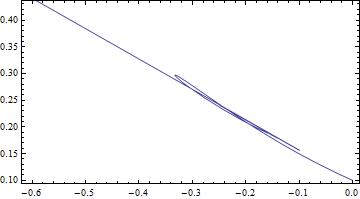}
\caption{Scale factor 2. $r\times10^{-1}$ against s. Interaction: $Q=q(3bH\rho_{m}+\gamma\dot{\rho}_{m})$}
\label{fig:rst3}
\end{figure}
\newpage
\section*{Discussion}
Two different types of interaction $Q=q(3bH\rho_{m}+\gamma\dot{\rho}_{m})$ and $Q=3bH\rho_{m}+\gamma\dot{\rho}_{m}$ between GUP Dark Energy and a fluid with a modified EoS were considered. In case of $Q=3bH\rho_{m}+\gamma\dot{\rho}_{m}$  we observe
phantom-quintessence-cosmological constant evolution for $\omega_{\small{tot}}$. Statefinder diagnostics reveal that we cross $\{ r=1, s=0\}$ twice corresponding to $\Lambda$CDM depending on $\beta$ parameter presenting in interaction $Q$, which indicates an impact of $\dot{\rho}_{m}$ in interaction. Departing from the section with positive $s$ and positive $r$ corresponding to the radiation phase of the Universe we cross $\{ r=1, s=0\}$ corresponding to $\Lambda$CDM and we back to radiation phase for late stage of the Universe (Fig. \ref{fig:rsLog}). Interacting by $Q=q(3bH\rho_{m}+\gamma\dot{\rho}_{m}),$ EoS $\omega_{\small{tot}}$ parameter has quintessence-phantom transition (Fig \ref{fig:omegatotLog2}). Here we are also able to cross
$\{ r=1, s=0\}$ corresponding to $\Lambda$CDM. In the second part of the article we considered a scale factor for which we have profile of $\ddot{a}$. We observe an interesting behavior for the model. For instance, $Q=3bH\rho_{m}+\beta\dot{\rho}_{m}$ interaction shows that when $p_{0}=\omega_{H}=\omega_{dH}=0$ and $\omega_{H2}=0$ during whole evolution of the Universe $\omega_{\small{tot}}=-1$. Then, with parameters $p_{0}=-1.5,~ \omega_{H}=-1~ \omega_{dH}=-0.5$ and $\omega_{H2}=-0.65,$ we saw, that $\omega_{\small{tot}}<=-1$ and it is true for whole evolution (Fig.\ref{fig:omegatot2}). In the case of positive $p_{0}=1.5$ we observe that in the early epoch $\omega_{\small{tot}}>0$ then there is a stage with $\omega_{\small{tot}}>-1$ indicating quintessence-like behavior. Finally for late time evolution $\omega_{\small{tot}}<=-1$ indicates phantom-like behavior (Fig.\ref{fig:omegatot2_1}). Statefinder diagnostics shows that the crossing of $\{ r=1, s=0\}$ is possible (Fig.\ref{fig:rs2}). Taking into account sign-changeable interaction between components of the mixture we observe that in case $p_{0}=\omega_{H}=\omega_{dH}=0$ and $\omega_{H2}=0$ EoS parameter $\omega_{\small{tot}}>=-1$ during evolution (Fig.\ref{fig:omegatot3_0}). For this case statefinder diagnostics presented in Fig.\ref{fig:rst3_0}. In case when parameters differ from $0$ we observe that with $p_{0}=1.5$ Universe starts with $\omega_{\small{tot}}>-1$, then continues its evolution and becomes $\omega_{\small{tot}}<-1$. Finally, for late time of the evolution $\omega_{\small{tot}}=-1$ (Fig.\ref{fig:omegatot3}). With $p_{0}=-1.5$ for EoS parameter we have Fig.\ref{fig:omegatot3_1}. For fixing real values of the parameters of the considered model, we should perform comparison with experimental data, which will be done in forthcoming articles. For all cases we see, that for late stages of the evolution mixture indicates itself as a cosmological constant. In an appendix we present results concerning to the Emergent and Intermediate scenarios for both cases of interactions.
\section*{Acknowledgments}
This research activity has been supported by EU fonds in the frame of the program FP7-Marie Curie Initial Training Network INDEX NO.289968.

\newpage
\section*{Appendix}

\begin{figure}[htp!]
\centering
\includegraphics[width=60mm]{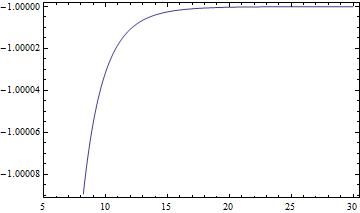}
\caption{Intermediate Scenario: $\omega_{\small{tot}}$ against t,
Interaction: $Q=3bH\rho_{m}+\gamma\dot{\rho}_{m}$}
\label{fig:omegatotinter}
\end{figure}

\begin{figure}[htp!]
\centering
\includegraphics[width=60mm]{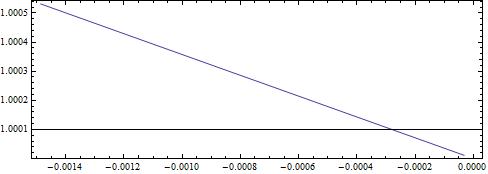}
\caption{Intermediate Scenario: r against s,
Interaction: $Q=3bH\rho_{m}+\gamma\dot{\rho}_{m}$}
\label{fig:rsinter}
\end{figure}

\begin{figure}[htp!]
\centering
\includegraphics[width=60mm]{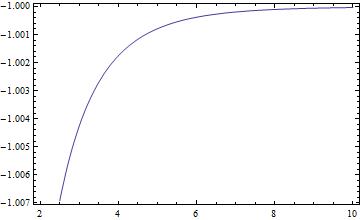}
\caption{Intermediate Scenario: $\omega_{\small{tot}}$ against t,
Interaction: $Q=q(3bH\rho_{m}+\gamma\dot{\rho}_{m})$}
\label{fig:omegatotinter}
\end{figure}

\begin{figure}[htp!]
\centering
\includegraphics[width=60mm]{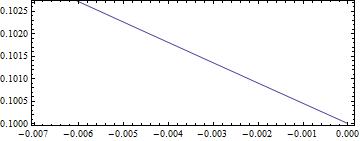}
\caption{Intermediate Scenario: r against s, Interaction: $Q=q(3bH\rho_{m}+\gamma\dot{\rho}_{m})$}
\label{fig:rsIntermediate2}
\end{figure}

\begin{figure}[htp!]
\centering
\includegraphics[width=60mm]{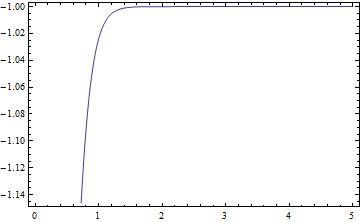}
\caption{Emergent Scenario: $\omega_{\small{tot}}$ against t,
Interaction: $Q=3bH\rho_{m}+\gamma\dot{\rho}_{m}$}
\label{fig:omegatotemerg}
\end{figure}

\begin{figure}[htp!]
\centering
\includegraphics[width=60mm]{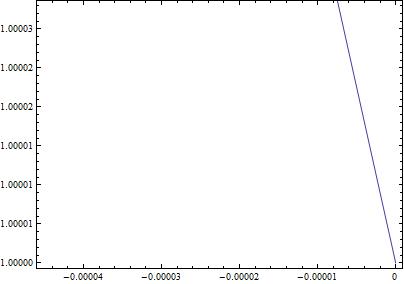}
\caption{Emergent Scenario: r against s,
$Q=3bH\rho_{m}+\gamma\dot{\rho}_{m}$}
\label{fig:rsemerg}
\end{figure}

\begin{figure}[htp!]
\centering
\includegraphics[width=60mm]{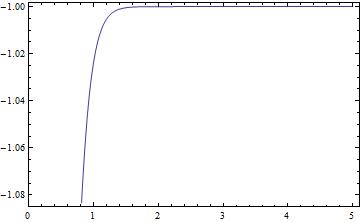}
\caption{Emergent Scenario: $\omega_{\small{tot}}$ against t,
Interaction: $Q=q(3bH\rho_{m}+\gamma\dot{\rho}_{m})$}
\label{fig:omegatotemerg2}
\end{figure}

\begin{figure}[htp!]
\centering
\includegraphics[width=60mm]{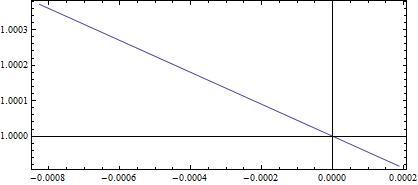}
\caption{Emergent Scenario: r against s,
$Q=q(3bH\rho_{m}+\gamma\dot{\rho}_{m})$}
\label{fig:rsemerg}
\end{figure}
\end{document}